\documentclass[nofootinbib,prd,aps,superscriptaddress,notitlepage]{revtex4-1}
\usepackage{hyperref,bm}
\usepackage{graphicx}
\usepackage{amsmath}
\usepackage[utf8]{inputenc}

\newcommand{\exv}[1]{\left\langle{#1}\right\rangle}

\newcommand{\ph}{\varphi}

\newlength{\szovszel}
\newlength{\slashszel} 
\newcommand*{\sls}[1]{\mbox{%
    \settowidth{\szovszel}{\ensuremath{#1}}%
    \settowidth{\slashszel}{\ensuremath{\slash}}%
    \hspace*{0.5\szovszel}%
    \hspace*{-0.5\slashszel}%
    \slash%
    \hspace*{-0.5\szovszel}%
    \hspace*{-0.5\slashszel}%
    \ensuremath{#1}%
  }}

\renewcommand{\d}{\partial}
\newcommand{\nn}{\nonumber\\}
\newcommand{\GeV}{\,\mathrm{GeV}}

\begin{document}

\title{Harmonic expansion of the effective potential in Functional
  Renormalization Group at finite chemical potential}

\author{G.G. Barnaf\"oldi}
\email{barnafoldi.gergely@wigner.mta.hu}
\affiliation{Wigner RSC, Budapest}
\author{A. Jakov\'ac}
\email{jakovac@phy.bme.hu}
\affiliation{Institute of Physics, Eotvos University, H-1117 Budapest, Hungary}
\author{P. P\'osfay}
\email{posfay.peter@wigner.mta.hu}
\affiliation{Wigner RSC, Budapest}
\affiliation{Institute of Physics, Eotvos University, H-1117 Budapest, Hungary}

\date{\today}

\begin{abstract}
  In this paper we propose a method to study the Functional
  Renormalization Group at finite chemical potential. The method
  consists of mapping the FRG equations within the Fermi surface into a
  differential equation defined on a rectangle with zero boundary
  conditions. To solve this equation we use an expansion of the
  potential in a harmonic basis. With this method we determined the
  phase diagram of a simple Yukawa-type model; as expected, the
  bosonic fluctuations decrease the strength of the transition.
\end{abstract}

\maketitle

\section{Introduction}
\label{sec:intro}

The Functional Renormalization Group (FRG) \cite{Gies:2006wv} is a
powerful nonperturbative method, it is nowadays a fair alternative to
other exact numerical methods, like the Monte Carlo lattice
simulations. Besides zero temperature applications \cite{Gies:2006wv}
there are efforts to adapt it to finite temperature and chemical
potential case
\cite{Herbst:2013ail,Drews:2013hha,Drews:2014wba,Drews:2014spa,Posfay:2015fpa,Eser:2015pka}. A
particularly interesting challenge is to reveal the equations of state
of the strong interaction, as this information could be directly used
in heavy ion collisions \cite{Herbst:2013ail} and in physics of
compact astrophysical objects \cite{Drews:2014spa}.

There are different formulation for the exact functional equation that
forms the basis of FRG computations, in this paper we will use
Wetterich equation \cite{Wetterich:1989xg,Wetterich:1992yh}. One can
also use different regulators \cite{Nandori:2012tc}, here we use
Litim's regulator \cite{Litim:2001up}. The Wetterich equation is
exact, but is valid in infinite dimensional operator space. In
practice one has to restrict the operator content to a finite set,
ie. we have to use an Ansatz for the effective action, and follow the
scale dependence only of terms present in this Ansatz. A popular
choice for bosonic systems is the Local Polynomial Approximation
(LPA), eventually extending it to wave function renormalization
effects (LPA'), for fermions restricting only to the renormalizable
operators (for fermions there exists also variants of the LPA method
\cite{Jakovac:2014lqa}).

The LPA (LPA') still provides a nonlinear partial differential
equation for the effective potential. For an analytic treatment, one
can power expand the effective potential, then we obtain a series of
ordinary differential equations for the scale dependence of the
coupling constants. With increasing number of terms taken into
account, one may hope that the solution converges to the exact
solution; as in some cases it can be really demonstrated
\cite{Mati:2014xma}.

Alternatively, one can discretize the potential,
representing the function with values taken at discrete points. This
again leads to a series of ordinary differential equations, and in
principle with raising the density of the representing points it will
reproduce the complete smooth function. Although this method seems to
be very promising, in practice it turns to be badly conditioned, very
sensitive to numerical errors \cite{Posfay:2015fpa}. There are therefore
variants of these methods to provide better convergence
\cite{Adams:1995cv,Fukushima:2010ji}.

At finite temperature and chemical potential the Bose-Einstein or
Fermi-Dirac distributions also appear in the FRG equations, which
makes the effective treatment of the LPA equations even more
tedious. Particularly at small temperatures and finite chemical
potential the Fermi-Dirac distribution becomes a near step function,
this makes the polynomial expansion useless, and the convergence
properties of the discretized version even worse.

Another unpleasant property of the discretized numerical method is
that the resulting infrared (IR) effective potential at zero scale
$k=0$ is strictly convex. At one hand it is a well known behavior of
the exact result obtained through Legendre transformation
\cite{Litim:2006nn}, but for field theoretical application the
coarse grained effective potential is much more comfortable to use, in
particular to identify a phase transition. The polynomial expansion
method, similarly to the conventional perturbation theory naturally
provides this potential, while in the discretized version the coarse
grained potential can be defined only at finite scale.

All this suggests that an analytic method similar to the polynomial
expansion working also at finite temperature and chemical potential
could be very useful. In this paper we propose such a method, first
studying the extreme case having zero temperature and finite fermionic
chemical potential. To demonstrate the way it is working we will use a
simple model with one bosonic and one fermionic degree of freedom,
coupled with Yukawa interaction. In order to find the scale dependent
effective potential $U(k,\ph)$, we can realize that the Fermi surface
splits the $k-\ph$ space into two parts, where the corresponding FRG
equations are different. One should solve the FRG equations at each
domain, and require that the solution is continuous at the border
line. The strategy we follow is to map the above problem to another
one where the Fermi surface is a rectangle, and we have zero boundary
conditions. Then we can use some function basis, in the present case a
harmonic basis, and solve the ordinary differential equation arising
for the coefficients. There are other possible choices for the
function base, recently in papers
\cite{Borchardt:2015rxa,Borchardt:2016xju,Borchardt:2016pif} 
the authors used Chebyshev polynomials.

We will discuss the convergence properties of our method, and find
out, how a coarse grained effective potential can be read off. With
this information we are able to identify the phase transition point as
well as explore the position of the borderline of the first and
second order phase transitions in the coupling constant space.

The structure of the paper is the following. After the Introduction,
in Section \ref{sec:model}, we introduce our model, write up the FRG
Ansatz, determine the FRG equations at finite temperature and chemical
potential, and at zero temperature we identify the Fermi-surface.
In Section \ref{sec:coordtrans} we discuss the coordinate
transformation needed to map the Fermi surface to a rectangle, leaving
the symmetries of the system intact. In Section \ref{sec:basis} we
propose a basis to handle the partial differential equation at hand,
and discuss the solution strategy. In Section \ref{sec:results} we
discuss the results: the effective potential at different
approximation level, the appearance of the first and second order
phase transition, and finally the order of the phase transition as a
function of the couplings, comparing it to the mean field analysis.
The paper is closed with a Conclusion section \ref{sec:concl}.

\section{The model and the FRG equation}
\label{sec:model}

We will use a simple Yukawa-type model with one bosonic and one
fermionic degree of freedom described by the bare action (defined at
scale $\Lambda$)
\begin{equation}
  \Gamma_\Lambda[\ph,\psi] = \int d^4x\left[\bar \psi(i\sls \d -
    g_0\ph)\psi + \frac12
    (\d_\mu\ph)^2-\frac{m_0^2}2\ph^2-\frac{\lambda_0}{24}\ph^4
  \right].
\end{equation}
This model has two phases, in the symmetric phase the fermion is
massless, in the spontaneously broken (SSB) phase the fermion mass is $g\exv{\ph}$.

As we want to treat this model with FRG, we need an Ansatz for the
effective action at scale $k$. Since in this paper the main goal is to
demonstrate the way how the finite chemical potential can be treated,
we choose the simplest possible Ansatz, where only the bosonic
effective potential depends on the scale:
\begin{equation}
  \Gamma_k[\ph,\psi] = \int d^4x\left[\bar \psi(i\sls \d -g\ph)\psi +
    \frac12 (\d_\mu\ph)^2 - U_k(\ph)\right].
\end{equation}
Although here neither wave function renormalization, nor the running
of the Higgs coupling are taken into account, both effects can be
easily adapted into the present method.

The Wetterich equation for this model reads
\begin{equation}
  \d_k U_k = \frac12\mathop{\mathrm{STr}} \ln (R_k + \Gamma_k^{(2)}),
\end{equation}
where STr means super-trace, $R_k$ is the regulator functional, and
$\Gamma_k^{(2)}$ is the second functional derivative of the effective
action. Using three-dimensional Litim's regulator, the corresponding
Wetterich equation reads at finite temperature $T$ and at finite chemical
potential $\mu$
\begin{equation}
  \d_k U_k = \frac{k^4}{12\pi^2}\left[
    \frac{1+2n_B(\omega_B)}{\omega_B} +
    4\frac{-1+n_F(\omega_F-\mu)+n_F(\omega_F+\mu)}{\omega_F}\right],
\end{equation}
where $n_B$ and $n_F$ are the Bose-Einstein and the Fermi-Dirac
distributions, respectively
\begin{equation}
  n_{B/F}(\omega)= \frac1{1\mp e^{-\beta\omega}},
\end{equation}
where $\beta=1/T$, while
\begin{equation}
  \omega_B^2 = k^2 +\d_\ph^2 U,\qquad \omega_F^2 = k^2 +g^2\ph^2.
\end{equation}
The initial condition of this equation is
\begin{equation}
  U_\Lambda(\ph) = \frac{m_0^2}2\ph^2+\frac{\lambda_0}{24}\ph^4.
\end{equation}

We want to discuss the zero temperature and at positive chemical
potential case. Then the Bose-distribution does not give
contribution, the Fermi-distribution reduces to
\begin{equation}
  n_F(\omega)\to \Theta(-\omega).
\end{equation}
Then the above equation simplifies to
\begin{equation}
  \d_k U_k = \frac{k^4}{12\pi^2}\left[\frac1{\omega_B} -
    4\frac{\Theta(\omega_F-\mu)}{\omega_F}\right].
\end{equation}
Note, that although seemingly this equation tells us that the fermion
distribution is active in the high energy case $\omega_F>\mu$, but in
fact here we just see the fermion vacuum fluctuations, while in the
low energy regime the vacuum fluctuations are compensated exactly by
the statistical fluctuations.

\begin{figure}[htbp]
  \centering
  \includegraphics[height=4cm]{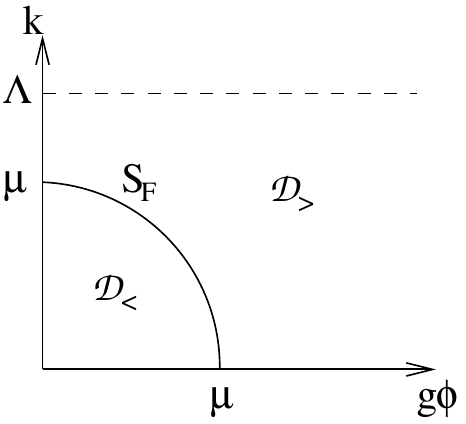}
  \caption{The structure of the differential equation. The $S_F$
    Fermi-surface is a circle in this coordinate system.}
  \label{fig:structure}
\end{figure}

The presence of the step function means that we have two different
domains, where two different differential equations evolve the
potential in $k$. The boundary of these domains is the
Fermi-surface $S_F$, it  can be determined from the equation
\begin{equation}
  \omega_F(k,\ph)\biggr|_{S_F} = \mu.
\end{equation}
The surface can be characterized either by $k=k_F(\ph)$ or by
$\ph=\ph_F(k)$. In our case these read
\begin{equation}
  k_F=\sqrt{\mu^2 - g^2\ph^2},\qquad \ph_F=\frac1g \sqrt{\mu^2-k^2}.
\end{equation}
The surface $S_F$, in terms of $k$ and $g\ph$, is a circle with radius
$\mu$; for $\mu=0$ it disappears. The Fermi-surface divides the
coordinate space into two parts; we will denote the high energy regime
by ${\cal D}_>$, the low energy regime by ${\cal D}_<$:
\begin{equation}
   {\cal D}_> = \left\{ (k,\ph) \,|\, k^2 +g^2\ph^2 > \mu^2\right\},\qquad
   {\cal D}_< = \left\{ (k,\ph) \,|\, k^2 +g^2\ph^2 < \mu^2\right\}.
\end{equation}
The structure of the differential equation is shown on
Fig.~\ref{fig:structure}.

In these domains the following differential equations hold:
\begin{subequations}
  \label{eq:Ueq}
  \begin{align}
    \label{eq:Ua}
  &\d_k U_k = \displaystyle\frac{k^4}{12\pi^2}\left[\frac1{\omega_B} -
    \frac{4}{\omega_F}\right],\hspace*{-3cm}
  &\mathrm{if}\; (k,\ph)\in {\cal D}_>,\\
    \label{eq:Ub}
  &\d_k U_k =
  \displaystyle\frac{k^4}{12\pi^2}\frac1{\omega_B},\hspace*{-3cm}
  &\mathrm{if}\; (k,\ph)\in {\cal D}_<.
  \end{align}
\end{subequations}
and the solution is continuous at $k=k_F$.

In a general problem we obtain equations with a similar structure,
although the Fermi surface which separates the ${\cal D}_>$ and ${\cal
  D}_<$ domains can have more complicated form. But we still have two
domains with different differential equations, and the requirement
that the two solutions are continuous at the border.

If the Fermi-surface can be described by a single-valued $k(\ph)$
function\footnote{This means that there is no turning back of the
  boundary line. If it is the case, we have to subdivide the ${\cal
    D}_<$ domain into parts for which the turning back can be
  avoided.}, then for all $(k,\ph)\in{\cal D}_>$ and for all $dk$
there is an open interval $I_{k+dk,\ph}$ centered around $(k+dk,\ph)$
for which $I_{k+dk,\ph}\subset {\cal D}_>$. Therefore we can compute
all the $\ph$ derivatives of the potential within this open interval,
which makes it possible to determine the potential at $(k,\ph)$. This
means that the solution on the ${\cal D}_>$ domain, starting from an
initial condition given at $k=\Lambda$, can be obtained without any
reference to the Fermi-surface. We can use the $\mu=0$ solution
there, which can be obtained using the standard FRG techniques, for
example with discretization or with polynomial expansion. The value of
the potential at the boundary can be determined by cutting out the
Fermi-surface from the zero chemical potential solution
\begin{equation}
  \label{eq:UbBCgen}
  V_0(k) = U_>(k,\ph_F(k)),
\end{equation}
where $U_>$ is the solution of \eqref{eq:Ua} with initial conditions
at $k=\Lambda$. Note that usually $\ph_F(k)$ can be determined only
after having the solution on ${\cal D}_>$, and usually it is a
complicated surface. In our model it means
\begin{equation}
  \label{eq:UbBC}
  V_0(k) = U_>(k,\frac1g\sqrt{\mu^2-k^2}).
\end{equation}

Discretization should work also in the ${\cal D}_<$ domain, starting
from the previously determined boundary conditions. To find a working
analytic method is a bigger challenge. The main problem is that the
boundary surface does not fit naturally to the coordinatization of the
potential, therefore a naive polynomial expansion does not work
here. There are two possible directions that we can follow. One is to
realize that the previous argumentation on the existence of a small
open interval remains true also here, but it should centered at
$(k-dk,\ph)$. Therefore we can solve the differential equation also on
${\cal D}_<$ without any reference to the Fermi surface, starting from
a general initial condition posed at $k=0$. The desired boundary
condition on the Fermi surface can then be phrased as a matching
condition for the initial condition. Although this approach is fully
sensible, by our experience, using a polynomial expansion for a
general initial condition, it is very badly conditioned, it very
easily leads to fast diverging solutions. Therefore a scan of initial
condition space is impossible, the convergent and divergent
regions are heavily mixed.

Trying out this possibility, we decided to follow another path,
introducing a new coordinate system which fits well to the Fermi
surface and the problem at hand. This will be discussed in the next
section.

\section{Coordinate transformation}
\label{sec:coordtrans}

A successful coordinatization must satisfy two requirements. The first
is that it should map the Fermi-surface to a rectangle, the second
requirement is that it should respect the symmetry of the differential
equation. In the present case the FRG equations are first order in
$k$, and second order in $\ph$. The coordinatization must not
introduce second order derivatives in both coordinates. Moreover we
have a $\ph\to-\ph$ symmetry. These requirements restrict the
possible transformations to
\begin{equation}
  (k,\ph)\mapsto (x,y),\qquad x=X(k),\; y=Y(k,\ph),\qquad
  U(k,\ph)=\tilde U(x,y),
\end{equation}
where $Y(k,-\ph)=-y$. Without restricting generality we can fix
$X(\mu)=0$ and $X(0)=\mu$. The inverse transformation is denoted as
\begin{equation}
  k= K(x),\qquad \ph = \Phi(x,y).
\end{equation}
We want that the image of the Fermi-surface is constant in $y$, let it
be $y=1$. This means in general
\begin{equation}
  Y(k, \ph_F(k) ) \equiv 1,
\end{equation}
This transformation maps the Fermi-surface to a rectangle with
$x\in[0,\mu]$ and $y\in[-1,1]$. The boundary conditions are
\begin{equation}
  \tilde U(x=0,y) = V_0(\mu),\quad \tilde U(x,y=\pm 1) = V_0(K(x))
\end{equation}

A simple realization of these constraints is the choice
\begin{equation}
  x = \ph_F(k),\quad y=\frac{\ph}{x},
\end{equation}
and so $\ph=xy$ and $K(x)=\ph_F^{-1}(x)$. The boundary is at $y=1$, so
the boundary conditions in these variables are
\begin{equation}
  \label{eq:boundaryxy}
  \tilde U(x=0,y) = V_0(\mu),\quad \tilde U(x,y=\pm 1) = V_0(K(x)) =
  U_>(K(x),x).
\end{equation}

Now we determine the form of the differential equation:
The derivatives can be written as
\begin{equation}
  \label{eq:ders}
  \d_k U = \frac{dX}{dk} \d_x\tilde U+ \frac{\d Y}{\d k} \d_y\tilde U =
   \frac{\ph'(k)}x\left( x \d_x\tilde U-y\d_y\tilde U\right),\qquad
   \d_\ph^2 U =  \frac1{x^2} \d_y^2\tilde U.
\end{equation}
Then the equation \eqref{eq:Ub} can be written as
\begin{equation}
  x\d_x\tilde U = y\d_y\tilde U  + \frac1{12\pi^2} \,\frac{k^4
    x^2}{\ph'_F}\, \frac1{\sqrt{(kx)^2+\d_y^2\tilde U}}\biggr|_{k\to K(x)}.
\end{equation}
We can also separate the boundary conditions from the solution by choosing
\begin{equation}
  \tilde U(x,y) =V_0(x) +\tilde u(x,y),
\end{equation}
then we have
\begin{equation}
  x\d_x\tilde u = -xV'_0 + y\d_y\tilde u  + \frac1{12\pi^2} \,\frac{k^4
    x^2}{\ph'_F}\, \frac1{\sqrt{(kx)^2+\d_y^2\tilde u}},
\end{equation}
where $k\to K(x)$, and the boundary conditions are
\begin{equation}
  \label{eq:tildeUeqshiftedincond}
  \tilde u(x=0,y) =\tilde u(x,y=\pm 1) =0.
\end{equation}
This is the generic form of the equation we should solve, appropriate
for any form of the Fermi-surface.

In our special case
\begin{equation}
  \label{eq:phiF}
  \ph_F = \frac1g \sqrt{\mu^2-k^2} =x,\qquad k=K(x) =
  \sqrt{\mu^2-g^2x^2} \qquad \ph_F' = -\frac{k}{g^2x}.
\end{equation}
After substituting back we find
\begin{equation}
  \label{eq:tildeUeqshifted}
  x\d_x\tilde u = -xV'_0 + y\d_y\tilde u  - \frac{g^2(kx)^3}{12\pi^2}
  \, \frac1{\sqrt{(kx)^2+\d_y^2\tilde u}}, 
\end{equation}
where $k\to K(x)$, and the boundary conditions remain $\tilde u(x=0,y)
=\tilde u(x,y=\pm 1) =0$.

\section{General solution with a complete system}
\label{sec:basis}

To solve eq. \eqref{eq:tildeUeqshifted} we try to expand the solution
$\tilde u(x,y)$ in terms of some basis. The polynomial expansion and
the discretization can be considered as a choice of basis, too. Here
we apply an orthonormal basis $h_n(y)$ with the properties
\begin{equation}
  \label{eq:basiscriteria}
  h_n(-y)= h_n(y),\qquad h_n(1)=0,\qquad \int\limits_0^1\!dy\,
  h_n(y)h_m(y) = \delta_{nm}.
\end{equation}
We will expand the desired solution in this base:
\begin{equation}
  \label{eq:expansion}
  \tilde u(x,y) = \sum\limits_{n=0}^\infty c_n(x) h_n(y).
\end{equation}
The boundary condition $\tilde u(x,y=\pm1)$ is automatically fulfilled
by the choice of the basis. The condition $\tilde u(x=0,y)=0$ requires
\begin{equation}
  \label{eq:incond}
  c_n(0)=0.
\end{equation}

We proceed by rewriting the partial differential equation
\eqref{eq:tildeUeqshifted} to an integro-differential equation. In
general if we have a partial differential equation of the form
\begin{equation}
  \d_x \tilde u = {\cal F}(x,y; \tilde u, \d_y\tilde u,\dots),
\end{equation}
then it is equivalent to a set of ordinary integro-differential
equations:
\begin{equation}
  c_n'(x) = \int\limits_0^1\!dy\, {\cal F}(x,y; \tilde u, \d_y\tilde
  u,\dots).
\end{equation}
We remark that this implies all FRG equations in a suitable coordinate
system, even at finite temperature and chemical potential. In our
special case we have
\begin{equation}
  \label{eq:integrodiff}
  x c'_n(x) = \int\limits_0^1\!dy\, h_n(y)\left[-xV'_0 + y
    \d_y\tilde u -
    \frac{g^2(kx)^3}{12\pi^2} \frac1{\sqrt{(kx)^2+\d_y^2\tilde u}}\right]
\end{equation}
with initial conditions \eqref{eq:incond}, where
\begin{equation}
  \d_y\tilde u =\sum\limits_{m=0}^\infty c_m(x) h'_m(y),\quad
  \d^2_y\tilde u =\sum\limits_{m=0}^\infty c_m(x) h''_m(y).
\end{equation}
This equation is exact, and treatable; in fact it represents an
alternative to the pointwise discretization.

But we can proceed and expand the inverse square root in the
last term around an arbitrary mass $M^2$:
\begin{equation}
  \label{eq:integrodiffexp}
  x c'_n(x) = \int\limits_0^1\!dy\, h_n(y)\left[-xV'_0 + y
    \d_y\tilde u - \frac{g^2(kx)^3}{12\pi^2} 
    \sum_{p=0}^\infty\left({-1/2}\atop p\right) \frac{(\d_y^2 \tilde u -
      M^2)^p}{\omega^{2p+1}}\right],
\end{equation}
where
\begin{equation}
  \omega^2 = (kx)^2 +M^2,\qquad
  \left(a\atop p\right) =
  \frac{a(a-1)\dots(a-p+1)}{p!}.
\end{equation}
The first few coefficients are $1, -\frac12,
\frac38,-\frac5{16},\frac{35}{128}$. This is still an adequate form
for a direct numerical integration, avoiding the problem of the
appearance of the small denominator in \eqref{eq:integrodiff}.

By expanding the numerator we encounter expressions like
$(\d_y^2\tilde u)^p$ which contains $p$-fold product of the second
derivative of the basis functions. These integrals can be performed
before the solution of the differential equation, since the
coefficients $c_n$ are functions of $x$, and so they do not influence
the $y$ integration. In the first two terms of
\eqref{eq:integrodiffexp} the integrals are explicit; let us denote
\begin{eqnarray}
  \label{eq:ABdef}
  && A_n = \int\limits_0^1\!dy\, h_n(y),\nn
  && B_{nm} = \int\limits_0^1\!dy\,h_m(y) yh'_m(y).
\end{eqnarray}
In the last term we need the quantity
\begin{equation}
   R^{(p)}_n(x) = \int\limits_0^1\!dy\, h_n(y)(\d_y^2  u(x,y))^p,
\end{equation}
which also means
\begin{equation}
  \label{eq:expandpowdiff}
  (\d_y^2  u(x,y))^p = \sum_{n=1}^\infty R^{(p)}_n(x) h_n(y).
\end{equation}
We can derive a recursion for $R^{(p)}_n(x)$. To that we will need the
auxiliary quantities $C_{nm}$ and $D_{nm\ell}$ defined as
\begin{eqnarray}
  \label{eq:CDdef} 
  && C_{nm} = \int\limits_0^1\!dy\, h_n(y)h''_m(y),\nn
  && D_{nm\ell} = \int\limits_0^1\!dy\, h_n(y)h_m(y)h_\ell(y).
\end{eqnarray}
Then
\begin{equation}
  R^{(1)}_n =\int\limits_0^1\!dy\, h_n \d_y^2u=
  \sum_{m=1}^\infty c_m \int\limits_0^1\!dy\,h_n h''_m =
  \sum_{m=1}^\infty C_{nm} c_m.
\end{equation}
Moreover, using \eqref{eq:expandpowdiff} we have
\begin{equation}
  (\d_y^2  u)^{p+1} = (\d_y^2  u)(\d_y^2  u)^p =
  \sum_{m\ell} R^{(1)}_mR^{(p)}_\ell h_m h_\ell = \sum_{nm\ell}
  R^{(1)}_mR^{(p)}_\ell D_{nm\ell} h_n, 
\end{equation}
which means
\begin{equation}
  R^{(p+1)}_n = \sum_{m\ell} D_{nm\ell}R^{(1)}_mR^{(p)}_\ell.
\end{equation}
In practice one uses symbolic manipulation program to perform this
recursion as a function of the $c_n$ coefficients. If we use ${\cal
  N}$ basis elements, then $R^{(p)}_n$ is a sum of ${\cal N}^p$ terms,
each of these is proportional to the product $c_{n_1}\dots
c_{n_p}$. It is evident, that after a certain order this expression
becomes untolerably long: then we should return to the explicit
integral equation \eqref{eq:integrodiffexp}, which is slower to solve,
but not too sensitive to the expansion order (and more stable, in
fact).

All in all, our equation \eqref{eq:integrodiffexp} can be written as
\begin{equation}
  \label{eq:integrodiffexpanded} 
  x c'_n(x) = -A_n xV'_0 + \sum\limits_{m=0}^\infty C_{nm} c_m(x) - 
    \frac{g^2(kx)^3}{12\pi^2} \sum_{p=0}^\infty\left({-1/2}\atop
      p\right) \frac1{\omega^{2p+1}} \sum_{r=1}^p \left(p\atop
      r\right) (- M^2)^{p-r} R^{(r)}_n.
\end{equation}
where $Q_n$ and $R^{(1)}_n$ depends linearly on $c_n$, and $R^{(p)}$
contains product of $p$ coefficients.

\subsection{Harmonic expansion}
\label{sec:harmexp}

A particularly stable algorithm can be based on the harmonic basis
functions. The basis is defined as
\begin{equation}
  h_n(y) = \sqrt{2} \cos q_n y,\qquad q_n=(2n+1)\frac\pi2,\qquad
  n=0,1,2\dots.
\end{equation}
This satisfies the criteria \eqref{eq:basiscriteria}. The coefficients
$A,\,B,\,C$ and $D$ can be worked out explicitly:
\begin{eqnarray}
  && A_n= \frac{\sqrt{2}(-1)^n}{q_n},\nn
  && B_{nm}= (-1)^{m+n}\frac{2q_nq_m}{q_m^2-q_n^2},\qquad
  B_{nn}=-\frac12,\nn
  && C_{nm} = -q_n^2 \delta_{nm}\nn
  && D_{nm\ell} = \frac{4\sqrt{2}(-1)^{m+n+\ell+1}q_mq_nq_\ell}
  {(q_m+q_n-q_\ell)(q_m-q_n+q_\ell)(-q_m+q_n+q_\ell)(q_m+q_n+q_\ell)}.
\end{eqnarray}

The form of the expansion coefficients suggest, why an expansion with
an orthonormal basis provides better convergence properties in the
solution of the FRG equations. Consider, for example, the expansion of
the term which is constant for $y<1$ and zero at $y=1$. In the harmonic
basis it has the expansion $\sum_n A_n h_n(y)$, here the coefficients
decrease as $\sim1/n$. We can also expand this function with
polynomials. There are different methods to do that, for example we can
require that the polynomial is zero at points $n/N$ where $n=0\dots
N-1$, and zero at $1$:
\begin{equation}
  U_{appr}(x) = 1-\frac{K(x)}{K(1)},\qquad\mathrm{where}\quad K(x) =
  \prod_{m=0}^{N-1} (x-\frac mN),
\end{equation}
because $K(\ell/N)=0$ for $\ell=1\dots N-1$.. The coefficient of the
highest power is $-1/K(1)$, its value is
\begin{equation}
  \frac1{K(1)}= \frac{N^N}{N!} \stackrel{N\to\infty}{\longrightarrow} e^N,
\end{equation}
using Stirling's formula $\ln N!=N\ln N-N+{\cal O}(\ln N)$. This means
that in the polynomial expansion the coefficient of the highest power
grows exponentially, which makes the numerical treatment of this
approximation more and more tedious.

\section{Results}
\label{sec:results}

Here we discuss the solution of \eqref{eq:integrodiffexpanded} in the
harmonic basis. We take ${\cal N}$ basis elements, and the expansion
of the inverse square root goes to power ${\cal P}$.

Before we start to discuss the solutions, we clarify a point in the
numerical treatment. We should start the evolution from $x=0$, but
then the derivatives have zero coefficient. To overcome this
difficulty we assume that for small $x$ the coefficients behave as
$c_n\sim x u_n$ (taking into account that $c_n(x=0)=0$
\eqref{eq:incond}). We then have in linear order
\begin{equation}
  x u_n = -xV'(0) A_n + x\sum_m u_mB_{mn} +{\cal O}(x^2),
\end{equation}
which yields
\begin{equation}
  \sum u_m (B_{mn}-\delta_{mn}) = V'(0).
\end{equation}
But, from \eqref{eq:UbBC} and from \eqref{eq:phiF}
\begin{equation}
  V'(x) = -\frac{g^2x}k \d_k U(k,x) + \d_\ph U(k,x)
  \stackrel{x\to0}{\longrightarrow} \d_\ph U(k,0) = 0,
\end{equation}
because of the parity of the potential. Therefore all the coefficients
$c_n(x)$ start at least as ${\cal O}(x^2)$. Numerically it is safe
therefore to start the evolution at $x_0\ll \mu$ with $c_n(x_0)=0$. In
practice we have taken $x_0/\mu \sim 10^{-4}$ as typical value.

\subsection{Mean field solution}
\label{sec:meanfield}

To test the reliability of the method, we apply it to an exactly
solvable case: the mean field (MF) approximation. Here we neglect the
bosonic fluctuations completely. Then what remains from equations
\eqref{eq:Ueq} is
\begin{equation}
  \d_k U_k = -\frac{k^4}{3\pi^2}\frac1{\omega_F}\Theta(k-k_F),
\end{equation}
with initial conditions at $k=\Lambda$. The equation is explicit, we
can solve it with a simple integration, resulting at
\begin{equation}
  \label{eq:rawMF}
  U(k,\ph)=
  U(\Lambda,\ph) +\frac{{{\cal N}_F}}{8\pi^2}\left[
    q \omega_F(q) \left(\frac23 q^2 -g^2\ph^2\right) + g^4\ph^4
    \ln(q+\omega_F(q)) \right]_{q=\max(k,k_F)}^{q=\Lambda}.
\end{equation}
The initial conditions and the radiative correction form the
renormalized potential
\begin{equation}
  U_{ren}(\ph) =  U(\Lambda,\ph) +\frac{{{\cal N}_F}}{12\pi^2}\left[ 
     \Lambda^4 - \Lambda^2 g^2\ph^2 + \frac32 g^4\ph^4
     \ln\frac{\Lambda}{M^*} + g^4\ph^4\left( \frac{3\ln 2}{2} - \frac
       78\right)\right],
\end{equation}
where assumed that $\Lambda$ is much larger than any other scale, and
we introduced the arbitrary $M^*$ mass, the renormalization
point. Assuming that $U_{ren}$ remains finite even for growing
$\Lambda$, the result can be made independent on the choice of the UV
cutoff. We write
\begin{equation}
  \label{eq:MFsol1}
  U(k,\ph) = U_{ren}(\ph) - \frac{{{\cal N}_F}}{8\pi^2}\left[ 
    q \omega_q \left(\frac23 q^2 -g^2\ph^2\right) + g^4\ph^4
    \ln\frac{q+\omega_q}{M^*} \right]_{q=\max(k,k_F)}.
\end{equation}

Taking into account that $\omega_F(k_F)=\mu$, the physical free energy
at finite chemical potential $U_\mu(\ph)=U(k=0,\ph)$, can be written as
\begin{equation}
  \label{eq:MFsol0}
  U_\mu(\ph) = U_0(\ph) - \frac{{{\cal N}_F}}{8\pi^2}\left[ 
    k_F \mu \left(\frac23 k_F^2 -g^2\ph^2\right) + g^4\ph^4
    \ln\frac{k_F+\mu}{M^*} \right]\Theta(\mu-g\ph).
\end{equation}

It is usual to represent the above result as an excess compared to the
zero chemical potential case. At zero chemical potential we have
\begin{equation}
  U_0(\ph) \equiv U(\mu=0,\ph) =  U_{ren}(\ph) - \frac{{{\cal
        N}_F}}{8\pi^2} g^4\ph^4 \ln\frac{g\ph}{M^*}.
\end{equation}
Using this formula we have
\begin{equation}
  U_\mu(\ph) = U_0(\ph) - \frac{{{\cal N}_F}}{8\pi^2}\left[ 
    k_F \mu \left(\frac23 k_F^2 -g^2\ph^2\right) + g^4\ph^4
    \ln\frac{k_F+\mu}{g\ph} \right]\Theta(\mu-g\ph).
\end{equation}
This corresponds to the usual one fermionic loop correction
\cite{Schmitt,Glendenning}.

Now let us solve the same problem with the method discussed in the
paper. The first point is to solve the equation at ${\cal D}_>$ at
$\mu=0$: this corresponds to the choice $k_F\to0$ in
\eqref{eq:MFsol1}. Next, we should cut off the Fermi surface, and
provide the initial conditions for ${\cal D}_<$. According to
\eqref{eq:boundaryxy} we have $V_0(x) = U_>(K(x),x)$, which is exactly
\eqref{eq:MFsol0} with $\ph\to x$:
\begin{equation}
  V_0(x) = U_\mu(x).
\end{equation}

The differential equation in ${\cal D}_<$ is not trivial even in the
absence of bosonic fluctuations. From \eqref{eq:tildeUeqshifted} we
should omit the last term, and have
\begin{equation}
  \label{eq:MFeq}
  x\d_x\tilde u = -xV'_0 + y\d_y\tilde u.
\end{equation}
It can be solved exactly, the solution is
\begin{equation}
  \tilde u(x,y) = V_0(xy) - V_0(x).
\end{equation}
The physical solution is
\begin{equation}
  U(k=0,\ph) = \tilde u(x=\frac\mu g,y= \frac{g\ph}\mu) +
  V_0(x=\frac\mu g) = V_0(\ph) = U_\mu(\ph),
\end{equation}
as we obtained earlier.

Since the equation \eqref{eq:MFeq} is similar to the fluctuating case,
the solution itself is again of similar form, the MF case is an
excellent testbed for verifying the reliability of the numerical
method.

Writing the potential in the explicit SSB form
\begin{equation}
  U(\ph) = \frac\lambda{24}(\ph^2-v^2)^2
\end{equation}
we have the relations among the couplings and the physical masses of
the fermion ($m_N$), the scalar ($m_\sigma$) and the pion decay
constant ($f_\pi$):
\begin{equation}
  \label{eq:couplings}
  v=f_\pi,\quad g = \frac{m_N}v,\quad \lambda = \frac{3m_\sigma^2}{v^2}.
\end{equation}
For the mean field case we used a nuclear-like parameters to test the
model with
\begin{equation}
  \label{eq:workingpoint}
  m_N=0.938\GeV,\quad f_\pi=0.093\GeV,\quad m_\sigma=m_N.
\end{equation}
We choose that chemical potential value, where the mean field
approximation predicts a first order phase transition (it is at
$\mu_{MF}\approx 0.6177 m_N$).

\begin{figure}[htbp]
  \centering
  \includegraphics[height=4cm]{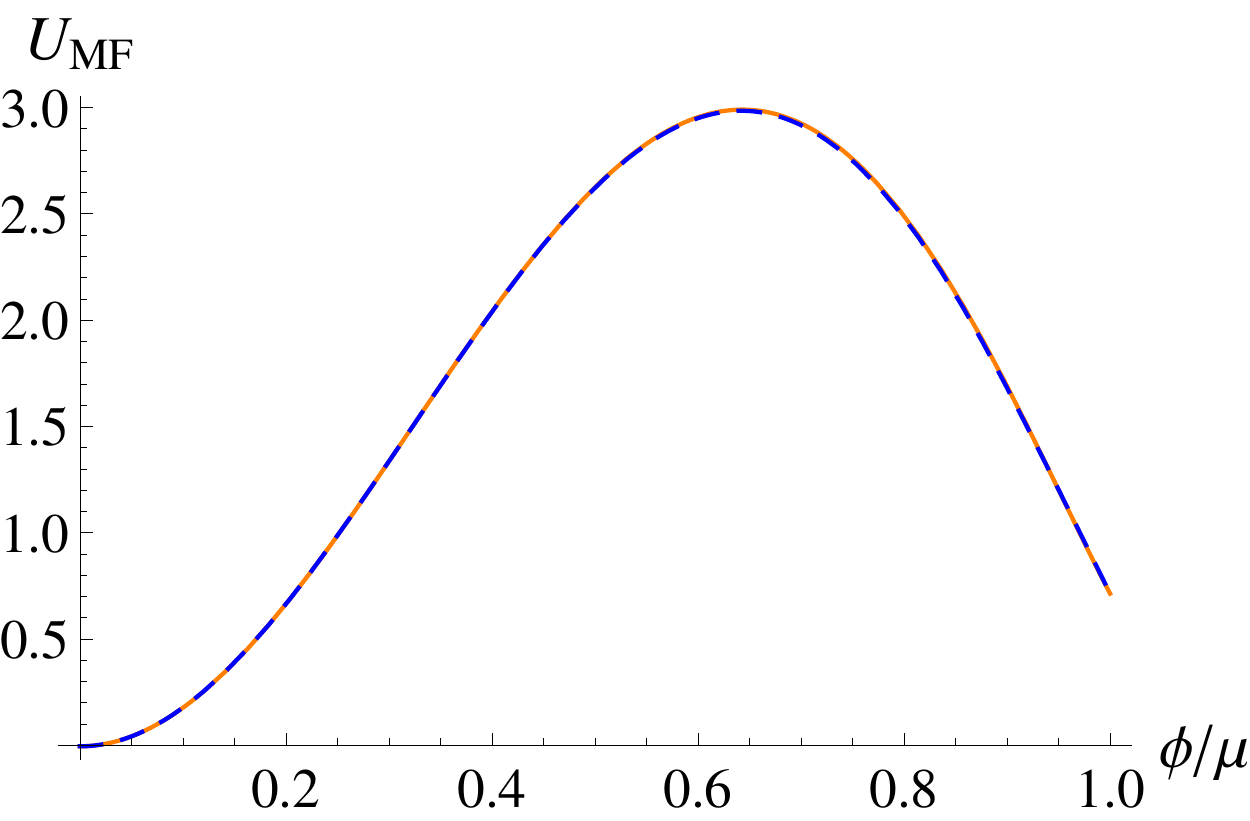}\qquad
  \includegraphics[height=4cm]{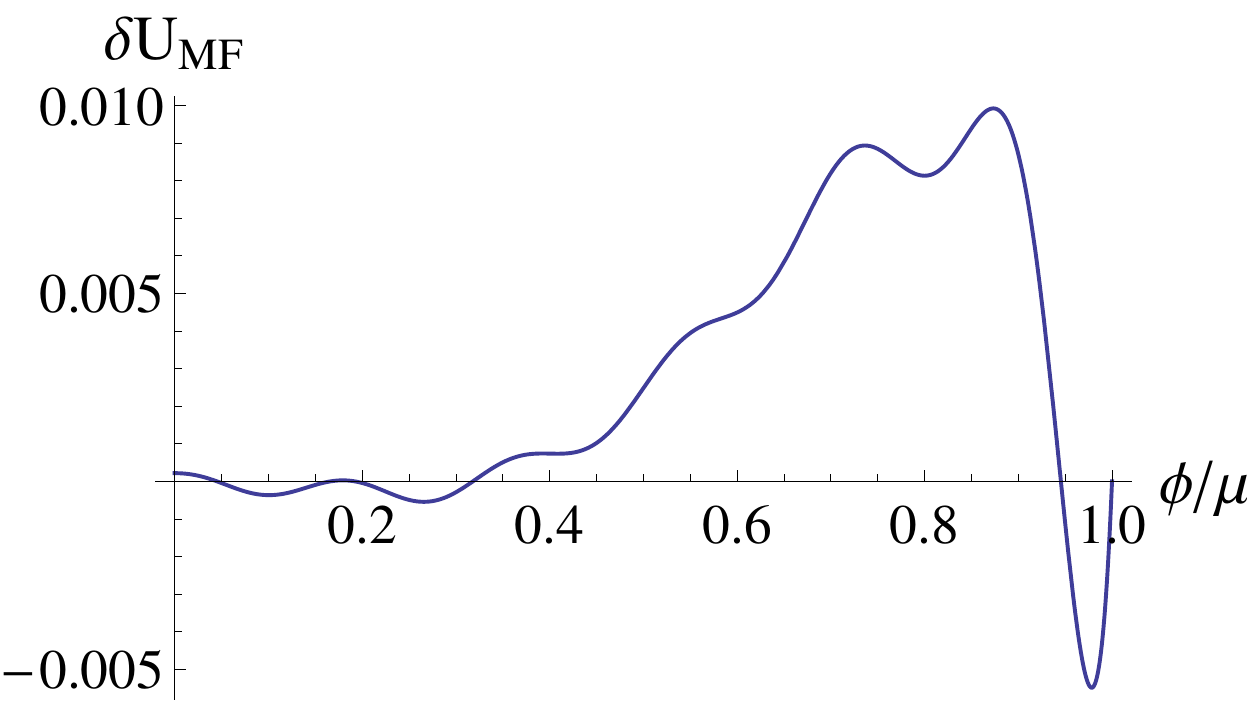}
  \caption{Reproduction of the exact mean field solution in a 10-element harmonic basis (left panel). The error of the reproduction (right panel).}
  \label{fig:MFrepro}
\end{figure}

If we have ${\cal N}=10$ basis elements, then the reproduction of the
mean field potential be seen on Fig.~\ref{fig:MFrepro}.
The error is on the percent level: this is the typical error in the
calculations also later on.

\subsection{Bosonic fluctuations}
\label{sec:bosfluct}

If we take into account the bosonic fluctuations, ie. we consider the
complete set of equations \eqref{eq:Ueq}, we should follow the
strategy described in Section \ref{sec:model}. First we solve the
$\mu=0$ case, ie. \eqref{eq:Ua} on the ${\cal D}_>$ domain. This
solution will be true at the outer domain at $\mu\neq0$, too. In this
domain the usual potential expansion approach works, and so we will
seek $U_0(k,\ph)$ as a polynomial in $\ph^2$, around an expansion
point $v^2$:
\begin{equation}
  U_k(\ph) = U_0 + \frac{m_k^2}2(\ph^2-v_k^2) +
  \frac{\lambda_k}{24}(\ph^2-v_k^2)^2.
\end{equation}
This form makes possible to treat the symmetric and the broken phase
together. In the former case we should expand around zero field value,
and so $v^2=0$; in the latter case $v$ denotes the minimum of the
potential, therefore $m^2=0$ must be chosen. But we can keep both
variable, and only in the course of the solution do we choose the
appropriate expansion point.

Expanding \eqref{eq:Ua} to power series in the variable $\ph^2$ around
$v^2$, the different orders must separately satisfy the equation. This
leads to the differential equations
($\omega_B^2=k^2+m^2+\frac\lambda3v^2$ and $\omega_F^2=k^2+g^2\ph^2$)
\begin{eqnarray}
  \d_k U_0-\frac{m^2}{2} \d_kv^2 =&& 
  \frac{k^4}{12\pi^2}\left[\frac1{\omega_B} - \frac4{\omega_F}\right]
  \nn
   \frac\lambda 6 \d_k v^2-\d_k m^2 =&& \frac{k^4}{12\pi^2}\left[
    \frac{\lambda}{2\omega_B^3}-\frac{4g^2}{\omega_F^3}\right]\nn
  \d_k\lambda =&& \frac{3k^4}{\pi^2}\left[\frac{\lambda^2}{16\omega_B^5}
    - \frac{g^4}{\omega_F^5}\right].
\end{eqnarray}

\begin{figure}[htbp]
  \centering
  \includegraphics[height=4cm]{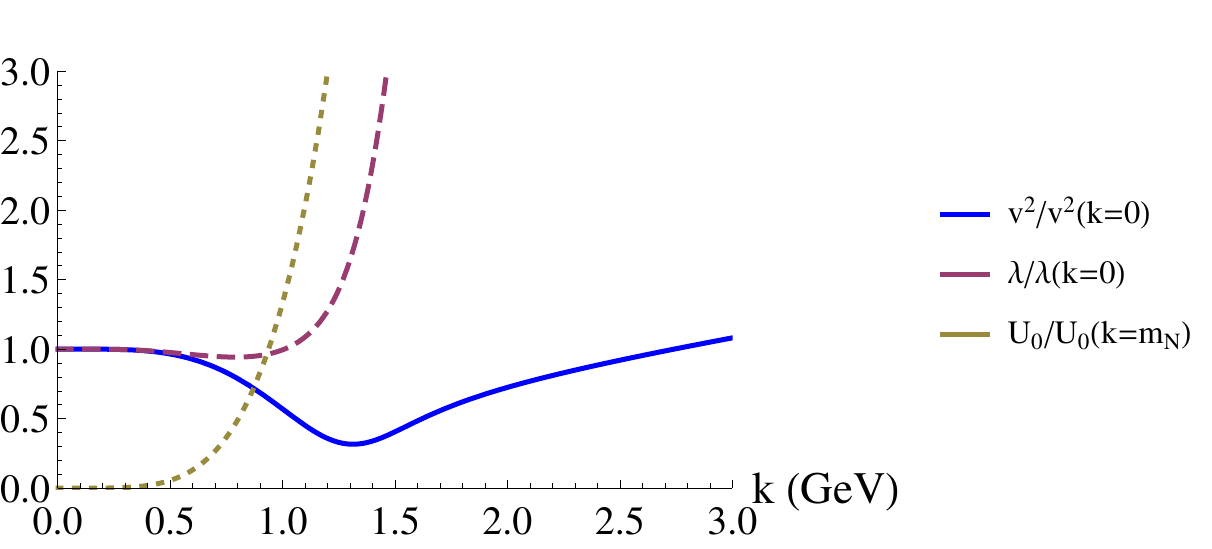}
  \caption{The running of the couplings in the ${\cal D}_>$ regime. In the plot the couplings are rescaled in order to show in a single plot.}
  \label{fig:couplingsrun}
\end{figure}

These equations can be solved. Using the working point
\eqref{eq:workingpoint} we remain in the symmetric phase. The low
energy part of the running of the couplings is shown in
Fig.~\ref{fig:couplingsrun}.

\begin{figure}[htbp]
  \centering
  \includegraphics[height=4cm]{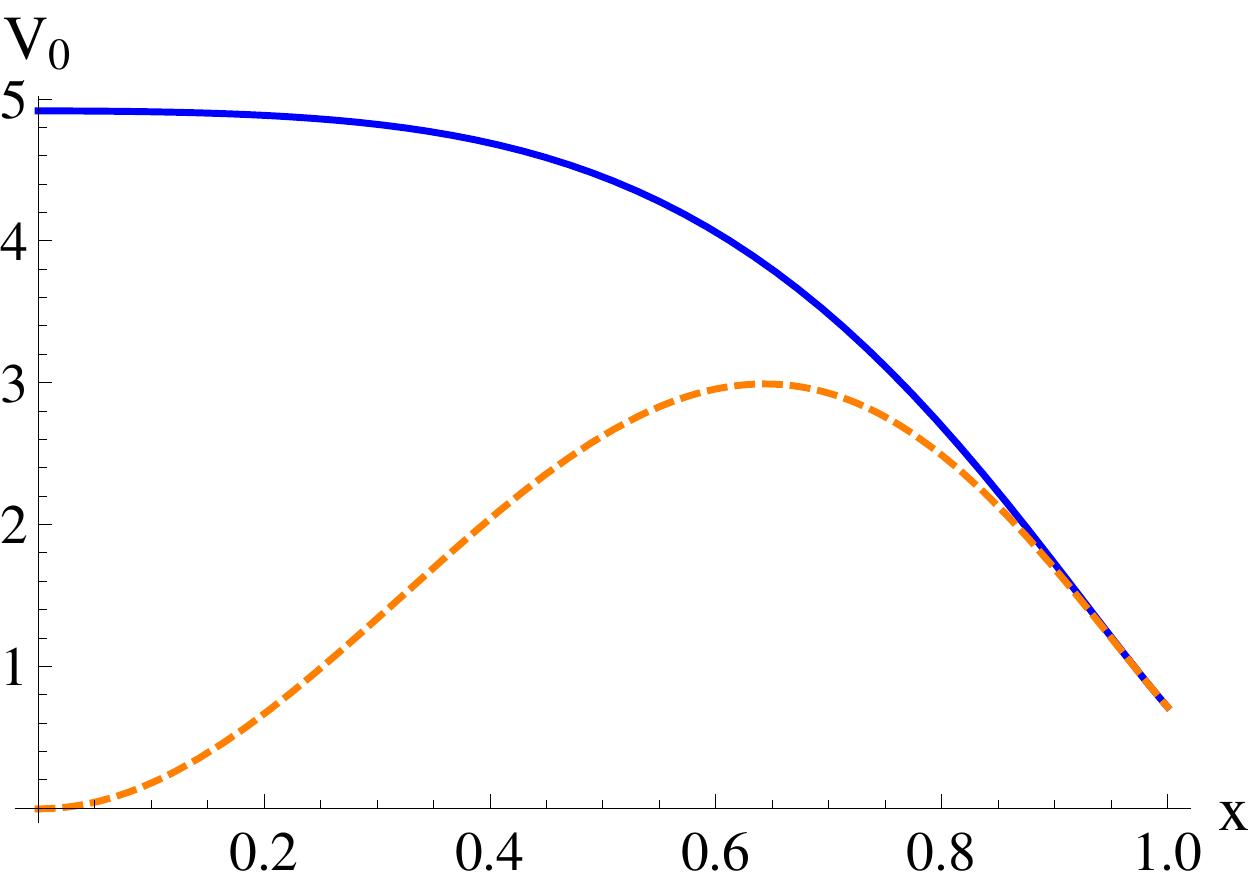}
  \caption{The boundary condition at $\mu=\mu_{MF}$ (solid blue
    line). To compare, we also plotted the mean field case (dashed
    orange line).}
  \label{fig:frgfinmu_boundarycond}
\end{figure}

Having the solution for all $k$ and $\ph$ we can cut off the boundary
conditions with the rule (cf. \eqref{eq:boundaryxy}) $V_0(x) =
U_>(K(x), x)$; this results in the figure
Fig.~\ref{fig:frgfinmu_boundarycond}.

\begin{figure}[htbp]
  \centering
  \includegraphics[height=4cm]{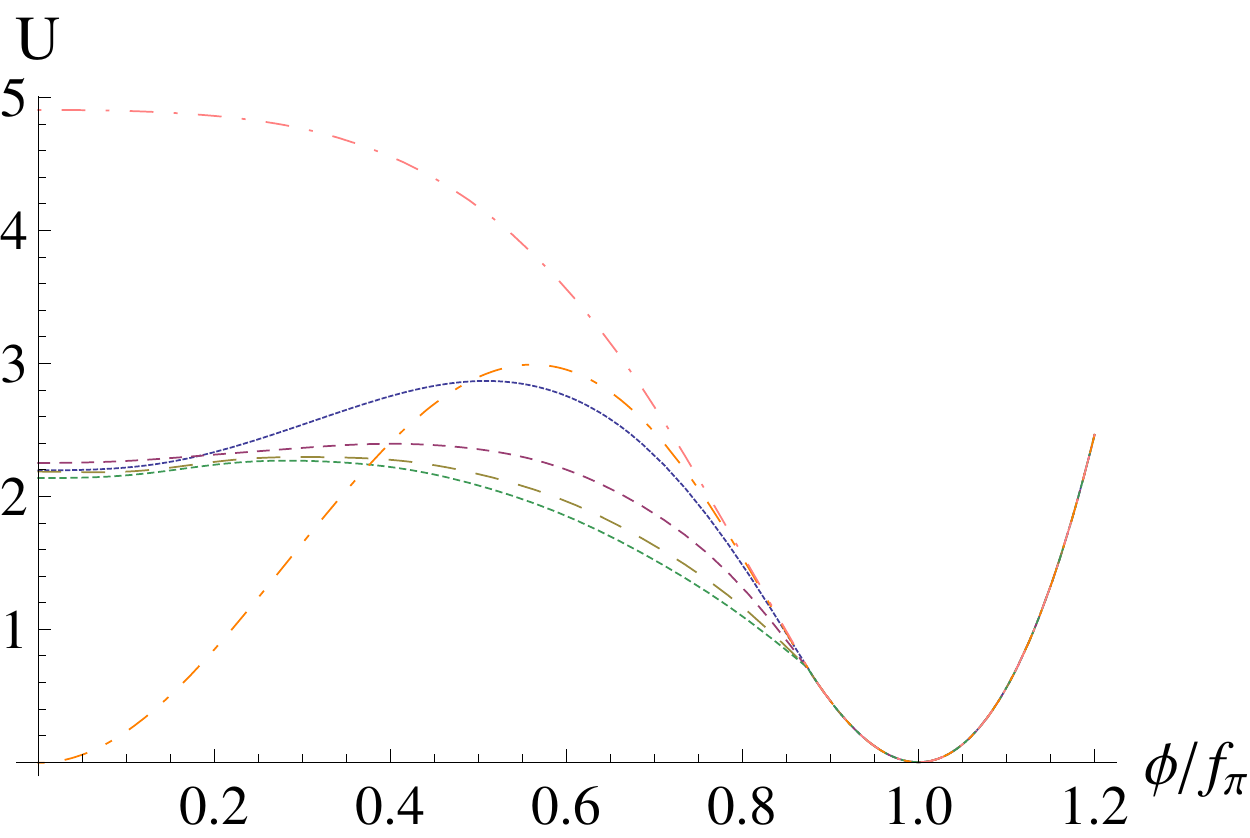}\hspace*{2em}
  \includegraphics[height=4cm]{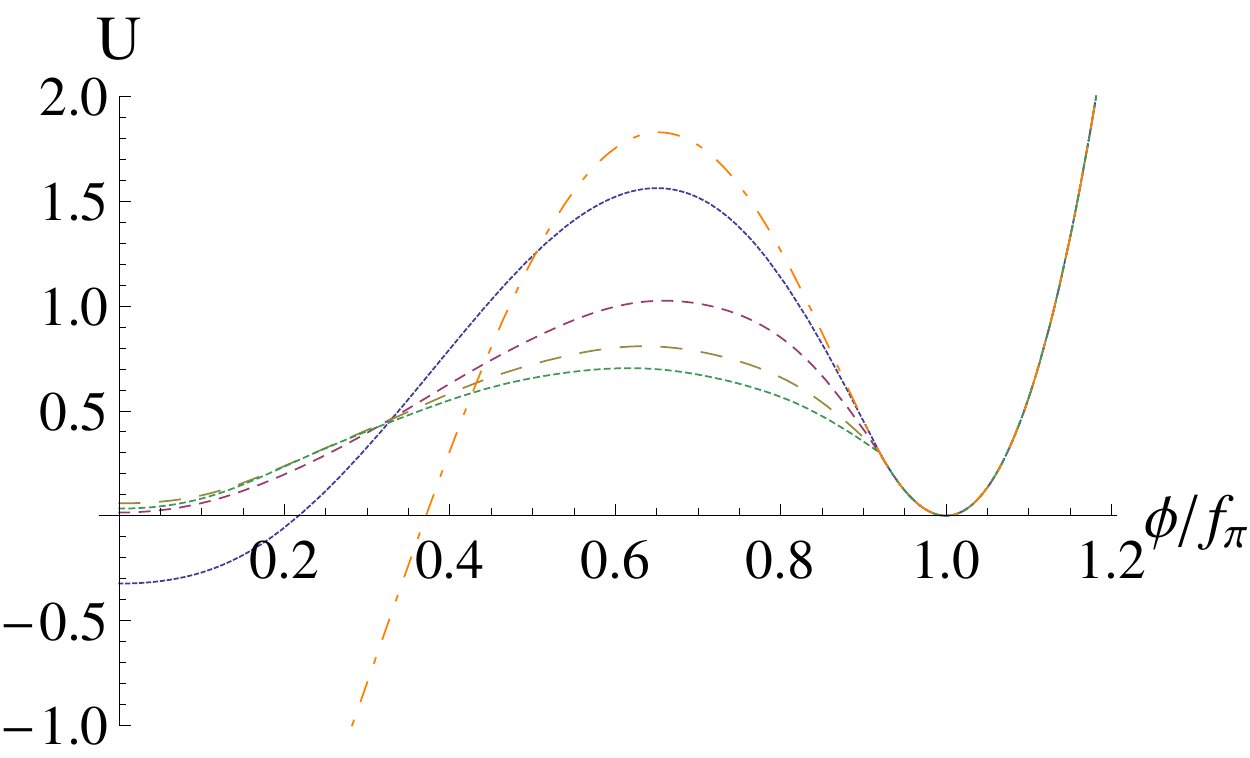}
  \caption{The effective potential resulting from the FRG
    calculation. The different curves correspond to different order of
    the expansion of the radiative correction in
    \eqref{eq:integrodiffexpanded}. In the left panel $\mu=\mu_{MF}$
    the mean field critical chemical potential, in the right panel
    $\mu=\mu_c=1.053\mu_{MF}$ the true critical chemical potential. In
  the right panel the ${\cal P}=0$ case is not shown, and the mean
  field curve goes down to $U_{MF}(\ph=0)=-2.92$.}
  \label{fig:frgfinmu_frg1}
\end{figure}

Starting from this boundary condition we can solve
\eqref{eq:integrodiffexpanded}. We used ${\cal N}=8$ basis elements,
and the inverse square root was expanded to ${\cal P}=0,1,2,3$ and $4$
order. The complete result depends on the choice of the regularizing
mass: we played around with its value to find the best convergence, in
the present case it was $U''(\ph=\mu)/18$. The resulting curves
can be seen on the right panel of Fig.~\ref{fig:frgfinmu_frg1}, where
also the result of mean field analysis is shown.
The same analysis at a somewhat larger chemical potential
($\mu_c=1.053\mu_{MF}$) results the plot on the left panel of
Fig.~\ref{fig:frgfinmu_frg1}: this is the critical chemical potential
of the first order phase transition belonging to the given couplings
\eqref{eq:couplings} describing the working point
\eqref{eq:workingpoint}.

To interpret what we can see in the figures, we recall that the exact
effective potential must be convex. Since the boundary condition fixes
the value of the potential at the border, if $\mu\le\mu_c$, then the
exact effective potential is a constant. So if ${\cal P}\to\infty$ we
should find shallower and shallower effective potential until it
flattens out completely. This means that we cannot expect pointlike
convergence.

In the figures of Fig.~\ref{fig:frgfinmu_frg1} one can identify two
regimes: at those $\ph$ values, where the curvature of the potential
is positive, we can observe rather good convergence. The change of the
effective potential from the mean field (${\cal P}=0$) to the one loop
(${\cal P}=1$) approximation is large yet, from ${\cal P}=1$ to
${\cal P}=2$ the change is moderate, but still observable (especially
on the right panel). But the potential values of ${\cal P}=2,3,4$ are
almost the same, there is a small deviation between them. 

In the other regime, where the curvature is negative, the effective
potential is concave, we find bad convergence. This can be expected,
since the exact result must have a non-negative curvature.

Based on these observations we can treat ${\cal P}=2$ or 3 as the best
approximation for the coarse grained effective potential, at least
when we interested in the thermodynamics. At those points, where the
exact effective potential has positive curvature, these curves already
converge to the good values; at those points, where the curvature is
negative, the value of the ${\cal P}=2$ or 3 appriximation
is irrelevant. We must be aware, however, that quantities like the
surface tension can not be reliably calculated from these
approximations.

This method provides a very effective and fast approach to
calculate the (relevant part of the) exact effective potential. The
numerical evaluation of the ${\cal P}=2$ or 3 formula is just seconds,
the convergence is very good, the results are rather stable against the
actual choice of the number of basis elements, or the exact value of
the expansion mass $M^2$.

\subsection{Phase diagram}
\label{sec:phasediag}

In the thermodynamics a particularly interesting question is the order
of the phase transition as a function of the coupling constants. One
can find a curve $g_c(\lambda)$ that is the border line between the
first and second order regimes in this model. The goal of this
subsection is to find this function.

To understand what happens, we pick up couplings $\lambda, g$, and
start from a low chemical potential, where the system is in the broken
phase (by construction). The minimum of the potential is at $\ph=\pm
v\neq0$ (because of the $Z_2$ symmetry $\ph\leftrightarrow-\ph$), the
curvature at $\ph=0$, denoted by $m_0^2$, is negative $m_0^2<0$.

\begin{figure}[htbp]
  \centering
  \includegraphics[height=4cm]{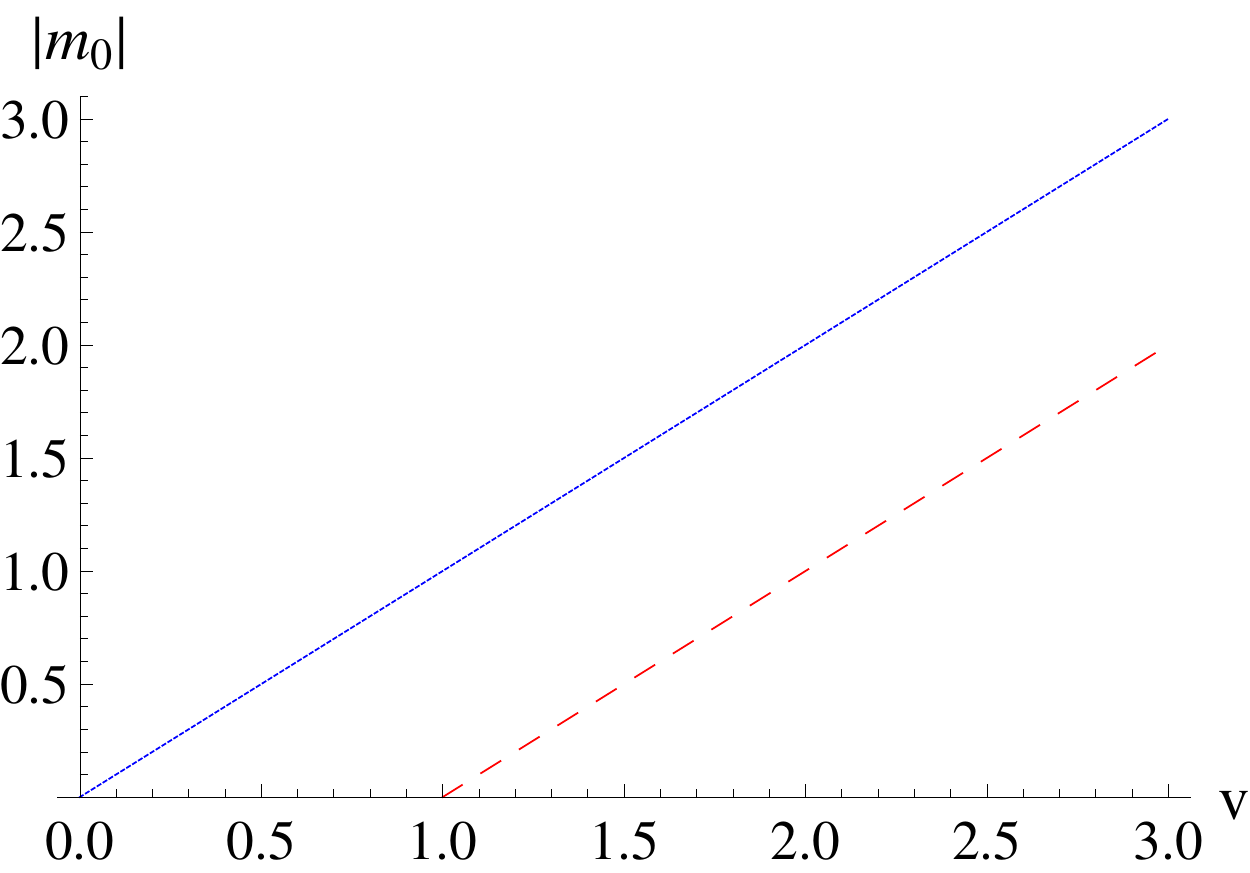}
  \caption{Identification of the order of the phase transition: if the
    $|m_0|$ vs. $v$ curve crosses zero at $v=0$, then we have a second
    order, if it crosses zero at positive $v$, a first order
    transition.}
  \label{fig:order}
\end{figure}

If we start to increase the chemical potential, both $v$ and $|m_0^2|$
decrease, sooner or later they will cross zero. There are two
possibilities: $|m_0^2|$ reaches zero before $|v|$, or they reach zero
in the same time. In this model the $Z_2$ symmetry requires zero slope
at $\ph=0$, therefore there is either a minimum or a maximum. For
$m_0^2<0$ it is a maximum, if $m_0^2=0$ while $v$ still positive, then
a new minimum is formed, therefore we have a first order transition
later. If $v$ and $m_0^2$ reach zero in the same time, then we have a
second order transition. Therefore we can identify the order of the
phase transition observing only the second order derivative at $\ph=0$
and the position of the minimum (cf. Fig.~\ref{fig:order}).
Note that in this way at most second order derivatives are needed for
the identification of the order of the transition.

\begin{figure}[htbp]
  \centering
  {\includegraphics[height=4.5cm]{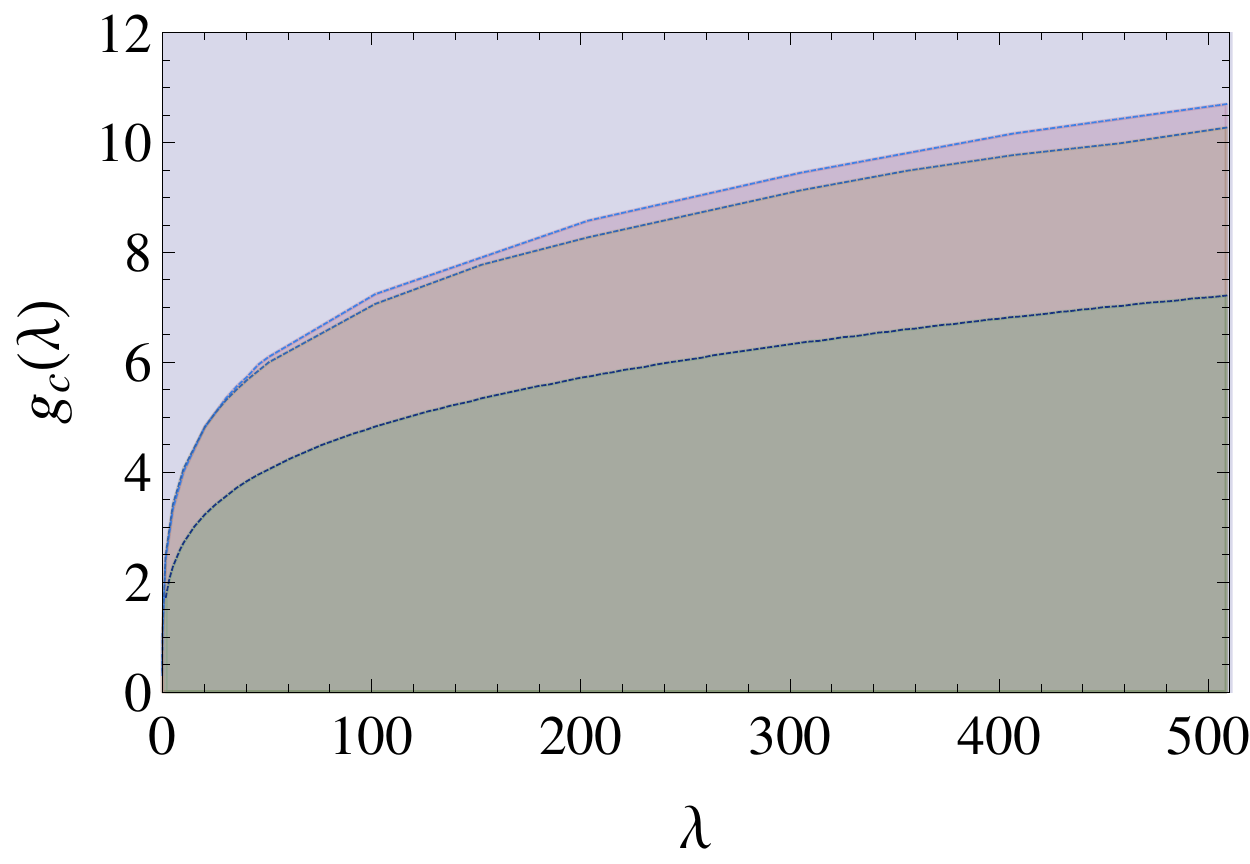}}
  \caption{Phase diagram in the coupling constant plane for the scalar
    Yukawa model. In the darker regimes we have second order, in the
    lighter regimes first order phase transition. The curves from
    bottom to top correspond to the mean field, one loop and FRG
    calculations, respectively. There are coupling pairs, where mean
    field predicts first order transition, but according to the FRG
    calculation it is still second order.}
  \label{fig:borderline}
\end{figure}

In the first order regime the $m_0=0$ point is reached at a certain
$\mu=\mu_u(\lambda,g)$ value, where $v=v_c(\lambda,g)$. We can solve
the $v_c(\lambda,g_c(\lambda))=0$ equation to determine the point,
where a second order transition first shows up. The resulting
$g_c(\lambda)$ curve is the borderline of the first and second order
regimes. This curve can be determined with the method discussed above
for the mean field and for the one loop computation and for the FRG
case. The one loop can be accessed from FRG as the first term in the
expansion of the inverse square root using the unimproved $\sigma$
mass as the expansion point. In the FRG computation we used a
12-element basis, and second and third order expansion of the inverse
square root. The resulting plot can be seen on
Fig.~\ref{fig:borderline}.
As we see, at small Yukawa couplings the phase transition is second
order, while for large Yukawa couplings it is first order. In the plot
the darker regions denote second, the lighter the first order
regime. The borderline of the two regimes depend on the order of the
approximation: the mean field calculation (ie. tree level
approximation for bosons) predicts the strongest phase transition, in
the sense that for increasing $g$ in this approximation comes the
first order regime earliest. So there is a regime, where the mean
field already predicts first order phase transition, but the exact
result is still a second order one. The one loop approximation is
already fairly good, it predicts only slightly stronger phase
transition than the exact result. These results are in accord with
observations in other models \cite{Herbst:2013ail,Jakovac:2003ar}.

We remark that all phase boundaries can be well fitted by a
\begin{equation}
  g_c(\lambda)\approx C\lambda^{1/4}
\end{equation}
analytic curve. The reason for this good fit is that if the effective
potential is analytic around $\ph=0$, then we can power expand
it around $\ph=v$ for small enough vacuum expectation value:
\begin{equation}
  U(\ph) = A(\ph^2-v^2)^2 + B(\ph^2-v^2)^3+\dots.
\end{equation}
If $A>0$ while $v\to0$ then it is a second order transition, if $A<0$
while $v\to0$, this describes a first order transition, at the border
line $A=0$. Perturbative arguments dictate an expansion in the
couplings as
\begin{equation}
  A = A_1\lambda +A_2\lambda^2 + A_3 \lambda^3 - A_4g^4 +\dots
\end{equation}
with some coefficients. This can be made vanish for small $\lambda$
and $g$ by the assumption $A_1\lambda\approx A_4g^4 + {\cal
  O}(g^6)$. This argumentation, although perturbative, seems to be
remain valid even in the domain of stronger couplings, too. The
coefficient in the $g_c(\lambda)\sim\lambda^{1/4}$ relation already
depends on the level of approximation, but converges nicely: the ratio
of the mean field and the exact result is about $\sim1.45$, for the
ration of the one loop and mean field results it is only $\sim1.04$.

\begin{figure}[htbp]
  \centering
  \includegraphics[height=4cm]{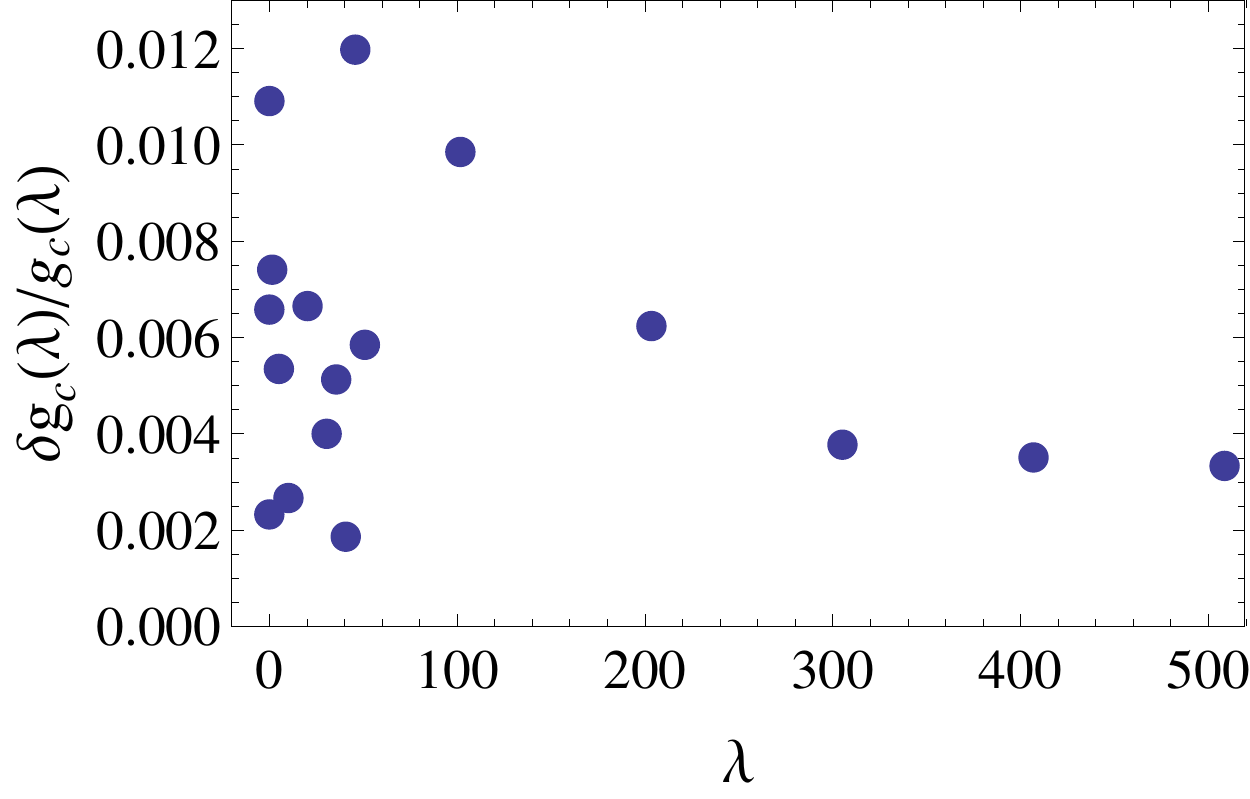}
  \caption{Relative difference of the border line using second or
    third order expansion of the inverse square root. The difference
    is below 1.5\%, which hints for a good convergence.}
  \label{fig:dg}
\end{figure}

Fig.~\ref{fig:dg} serves to demonstrate the convergence at larger
orders. As we see, the relative deviation of the the second and third
order expansion of the square root is $0.2$-$1.2\%$, these yields
curves within line width in Fig.~\ref{fig:borderline}. This supports
also our earlier finding, that in the physically sensible regime the
second order expansion of the inverse square root is already close
enough to the exact one, provided we use an appropriate expansion
point.

\section{Conclusions}
\label{sec:concl}

In this paper we proposed a systematic approximation scheme for the
treatment of the FRG equations at finite chemical potential. In short,
the Fermi surface divides the available $(k,\ph)$ domain into two
parts: in the high energy ${\cal D}_>$ domain the vacuum equations are
valid, while in the small energy ${\cal D}_<$ regime the statistical
fluctuations of the fermionic modes are also present. Starting from
initial condition posed at $k=\Lambda$, in the ${\cal D}_>$ domain the
solution is not affected by the presence of the Fermi surface. This
solution, on the other hand, provides a boundary condition for the
equation in the ${\cal D}_<$ domain. To be able to solve the equations
in the ${\cal D}_<$ domain, we introduced new variables ($x$ and $y$),
with them we transformed the Fermi-surface into a rectangle. After
separating the boundary conditions we had to solve an inhomogeneous
nonlinear partial differential equation in this rectangle with zero
boundary conditions. We looked for the solution in a form of a series
in a complete basis of harmonic functions, then for the coefficients
we had a set of ordinary differential equations. We performed another
expansion, namely we expanded the $\sim(k^2+u'')^{-1/2}$ type term in
powers of $(u''-M^2)^n$, where $M^2$ was an appropriately chosen
mass.

Having two expansions (the number of basis elements, and the order of
the expansion of the inverse square root), we had to study the
convergence in both. As it turned out, a moderate number of basis
elements (6-12) is enough to have a good convergence. The order of the
expansion of the square root is a more delicate question. As we have
argued in the paper, the exact effective potential should be convex,
unlike the  physically more useful coarse grained effective
potential. What we really need is the convergence in the regimes,
where the potential is convex, in the unphysical, concave parts we can
not expect good convergence properties. The convergence in the convex
part of the effective potential can be achieved (by a proper choice
of the expansion point $M^2$), already by taking into account a second
or third order expansion of the inverse square root (corresponding to
a linear or quadratic power of $(u''-M^2)$).

This method is a semi-analytic approach to the solution of the FRG
equations at finite chemical potential, similar to the polynomial
expansion used at zero chemical potential. It is a powerful and
accurate method, the differential equations for the evolution of the
coefficients in the harmonic basis are well conditioned (unlike in the
expansion in polynomial basis), so the numerical treatment is easily
accessible and fast.

With this method we determined the phase structure of the Yukawa
model, ie. we determined the line in the coupling constant plane,
where the boundary between the first and second order phase transitions
is situated. One can compare the results of the different
approximations (mean field, one loop, exact). The main result from
this study is that the bosonic fluctuations soften the strength of the
transition, making it more ``second-order-like''. More precisely,
there is a region in the coupling constant plane, where the mean field
calculation predicts a first order transition, but in fact it is a
second order transition according to the exact result. In this sense
the mean field predicts the strongest phase transition, then the one
loop approximation follows, the exact result gives the weakest
transition.

As a future prospect, this method can be applied to other models where
the small temperature, finite chemical potential regime is physically
significant. Moreover, as we also mentioned in the paper, there is a
natural expansion towards the finite temperature and finite chemical
potential regime.

\section*{Acknowledgments}

The author thanks for instructive discussions with A. Patk\'os,
Zs. Sz\'ep and G. Mark\'o. This work is supported by the Hungarian
Research Fund (OTKA) under contract No. K104292.

\end{document}